# Dzyaloshinskii-Moriya interaction chirality reversal with ferromagnetic thickness

Capucine Gueneau[1,¶], Fatima Ibrahim[1,¶], Johanna Fischer[1,*], Libor Vojáček[1], Charles-Élie Fillion[1], Stefania Pizzini[2], Laurent Ranno[2], Isabelle Joumard[1], Stéphane Auffret[1], Jérôme Faure-Vincent[1], Claire Baraduc[1], Mairbek Chshiev[1,3], and Hélène Béa[1,3,†]

[1]Univ. Grenoble Alpes, CEA, CNRS, Spintec, 38000 Grenoble, France

[2]Univ. Grenoble Alpes, CNRS, Néel Institute, 38042 Grenoble, France

[3]Institut Universitaire de France (IUF), 75000 Paris, France

¶ C.G. and F.I. contributed equally

**ABSTRACT**. In ultrathin ferromagnetic films sandwiched between two distinct heavy metal layers or between a heavy metal and an oxide layer, the Dzyaloshinskii-Moriya interaction (DMI) is of interfacial origin. Its chirality and strength are determined by the properties of the adjacent heavy metals and the degree of oxidation at the interfaces. Here, we demonstrate that the DMI chirality can change solely with variations in the thickness of the ferromagnetic layer - an effect that has not been experimentally studied in details or explained until now. Our experimental observation in the trilayer system Ta/FeCoB/TaOx is supported by *ab initio* calculations: they reveal that variations in orbital filling and inter-atomic distances at the interface, driven by the structural relaxations in the ultrathin regime, lead to an inversion of DMI chirality. We hence propose a new degree of freedom to tune DMI chirality and the associated chiral spin textures by tailoring crystal structure e.g. using strain or surface acoustic waves.

## I. INTRODUCTION.

The Dzyaloshinskii-Moriya interaction (DMI) is an antisymmetric exchange interaction that can stabilize chiral spin textures or a cycloidal order. Recent experiments focused on ultrathin multilayers with perpendicular magnetic anisotropy (PMA), where spin textures such as skyrmions or chiral domain walls are stabilized at room temperature [1], and efficiently moved by injecting an electric current via spin-orbit torques [2,3]. Ultrathin heavy metal/ferromagnet/oxide (HM/FM/MOx) trilayers with PMA [4,5] offer a promising platform to study interfacial DMI which arises from the combination of spin-orbit coupling and structural inversion asymmetry at the two FM interfaces [6,7]. Specifically, the antisymmetric exchange between neighboring spins is mediated by the conduction electrons, which experience either spin-orbit interaction within the HM layer (Fert-Levy mechanism) [8] or a Rashba field at the FM/MOx interface (Rashba-type DMI) [9]. The strength and chirality of the interfacial DMI is quantified by a coefficient $D$ that depend on the nature of the HM and on the oxidation state of the oxide layer. It imposes a given domain wall type and chirality and is usually considered to be the sum of the contributions from the top and bottom interfaces. The interfacial character is modeled by $D = \frac{D_s}{t}$. Here, $t$ denotes the FM thickness, and $D_s$ the interfacial DMI coefficient comprising both interface contributions. The material-dependence of DMI in heavy-metal/ferromagnet systems has been demonstrated experimentally [10–13] and analyzed theoretically [14,15]. Furthermore, tuning the DMI chirality with the oxidation in FM/MOx interfaces has been predicted [16,17] and observed in similar trilayer structures [18]. However, the dependence of the DMI chirality solely on the ferromagnetic thickness without a variation in oxidation state has neither been systematically studied nor theoretically predicted.

Here, we present the DMI chirality inversion with ferromagnetic thickness, by both theoretical modeling (Fig. 1(a)) and experimental validation (Fig. 1(b)), to investigate the interfacial DMI in a weak DMI system, namely Ta/FeCoB/TaOx [19,20]. Perpendicular wedges of the ferromagnetic and top metal layers enable gradual variation of ferromagnet thickness independent of the FM/oxide interface oxidation gradient. This crossed double wedge method, unlike that in [18], differentiates interfacial effects from ferromagnetic thickness-related DMI changes, providing deeper insight into their respective contributions. The DMI chirality is unambiguously determined from the current-driven direction of domain walls (DWs) and skyrmions motion [21] due to spin-orbit torque, that depends on their clockwise (CW) or counter-clockwise (CCW) chirality (Fig. 1). Quantitative measurement of DMI coefficient is not accessible by other techniques in this weak DMI regime. To elucidate the underlying mechanisms of the observed DMI chirality changes, we performed *ab initio* calculations. The agreement between experiment and theory on the DMI chirality with both the ferromagnet thickness and oxidation state enables to propose unconventional mechanisms governing the DMI chirality. In particular, changes in electronic orbital filling due to modified inter-layer distance (z) between the FM and MOx layers can induce a DMI sign inversion with increasing the number of FM monolayers (Fig. 1(b)). This study paves the way to deterministic control of skyrmions and chiral domain walls by tuning the ferromagnetic thickness.

---

* contact author: Johanna.fischer@cea.fr

† contact author: Helene.bea@cea.fr

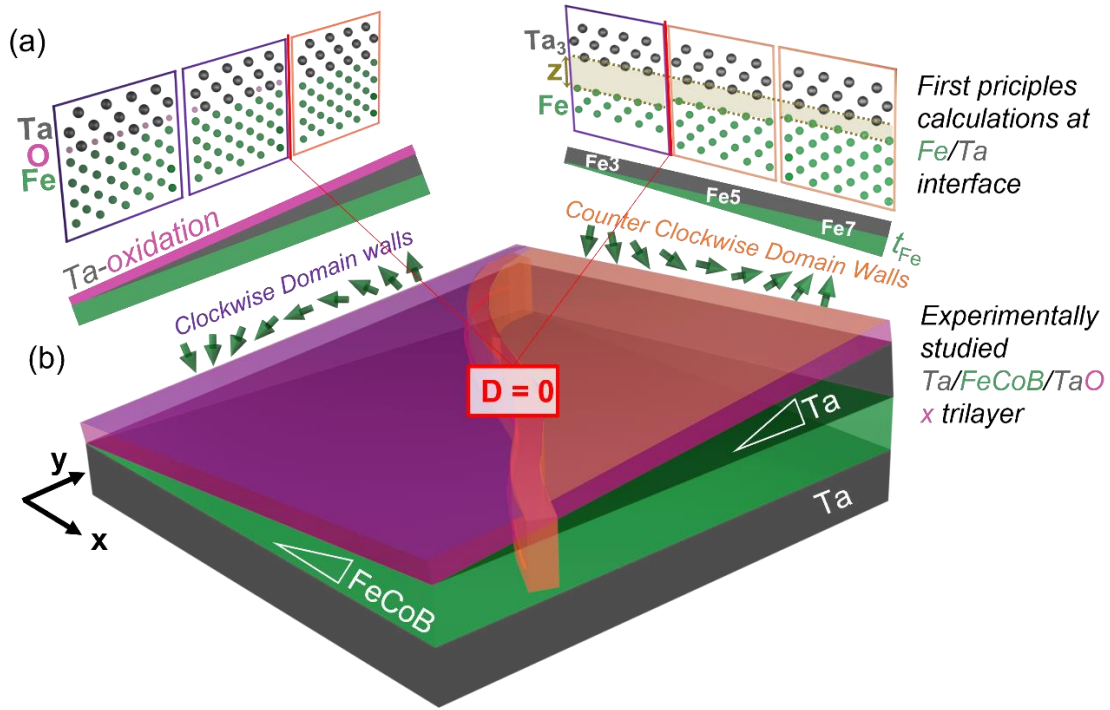

*Fig. 1 Experimentally studied (Ta/FeCoB/TaOx) double wedge (b) and theoretically analyzed Fe/Ta(Ox) top interface (a). The red band represents the DMI sign change (D = 0) with increasing ferromagnet thickness along x and decreasing degree of oxidation of the top heavy metal layer along y. Domain wall chirality (CW/CCW) is given by the violet/orange zones around D = 0, respectively. A projection (a) of the* ab initio *Fe/Ta(Ox) crystallographic structure at the top interface, is depicted along each wedge; green/gray/purple balls represent Fe/Ta/O atoms. z is the relaxed Fe-Ta plane distance.*

## II. METHODS

### A. Sample preparation

The sample was grown using on-axis and off-axis magnetron sputtering on a Si/SiO$_2$ 100 mm-diameter substrate. On-axis sputtering was employed to achieve uniform deposition, specifically for the bottom Ta layer. The target and the substrate are parallel and coaxial. A variation of 2 to 5 % thickness can be observed on the edge of the sample. We use off-axis deposition technique translating the sample with respect to the target, both remaining parallel, resulting in non-uniform thickness along one axis of the sample, following circles on the wafer (see Supplementary 1 for more details on the thickness variations and reconstruction). This technique was employed to obtain the two gradient layers of FeCoB and top Ta. We varied the deposition axis for each layer in order to have one gradient perpendicular to the other. The

*Contact author: johanna.fischer@cea.fr

†Contact author: helene.bea@cea.fr

FeCoB target composition is Fe72Co8B20. We reconstructed the map and take the convention that the FeCoB thickness varies along the x-axis meanwhile the top Ta thickness is along the y-axis. The x and y axes are thus not exactly fitting the horizontal and vertical directions on the wafer, but following the FeCoB and Ta thickness gradients. The presented maps include this reconstruction. The top Ta layer is oxidized by exposure to oxygen in a treatment chamber with a partial pressure of oxygen equal to the one of atmosphere. This leads to different oxidation states at the top interface named under-oxidized (FeCoB/Ta/TaOx), optimally oxidized (FeCoB/TaOx), and over-oxidized (FeCoBOx/TaOx). To protect the sample from extra oxidation and maintain its magnetic properties over time, a capping layer of 0.5 nm of Al in deposited *in-situ*. An annealing step is performed afterwards under vacuum at 225°C for 30 min. It increases the interface quality and enlarges the PMA region.

### B. *Ab initio* calculation method

Our first-principles calculations are based on the projector-augmented wave (PAW) method [22] as implemented in the VASP package [23–25] using the generalized gradient approximation [26] and including spin-orbit coupling. The Fe/TaOx interface is modelled by a supercell of three or five Ta monolayers on top of Fe with varied thickness from three to nine monolayers followed by a sufficient vacuum layer of 20 Å. A pure Fe layer is considered since our FeCoB target composition is Fe rich (Fe72Co8B20) and boron migrates away from the ferromagnetic layer during annealing. The TaOx wedge is modelled by either varying the Ta to O ratio in the Ta interfacial layer or changing the distance of the first O atoms from the Fe/Ta interface. The atomic coordinates are relaxed in the out-of-plane direction, with the in-plane parameter fixed to the bulk Fe, until the forces became smaller than 1 meV/Å. The DMI is calculated employing the constrained spin-spiral supercell method [27]. In this case, the DMI energy $E_{DMI}$ is defined as the energy difference of the clockwise (CW) and counterclockwise (CCW) spin configurations where $E_{DMI} = E_{CW} - E_{CCW}$. Fig. 2 demonstrates the corresponding supercell structures with spin spirals rotating in the $(xz)$-plane. Note that in this convention, a positive (negative) $E_{DMI}$ corresponds to a CCW (CW) DMI, respectively. The energy convergence criterium is set to $10^{-7}$ eV to ensure a good convergence of the DMI energies. A kinetic energy cutoff of 500 eV has been used for the plane-wave basis set and a Γ-centered $25 \times 25 \times 1$ k-mesh to sample the first Brillouin zone while a $6 \times 24 \times 1$ mesh is used for the DMI calculations.

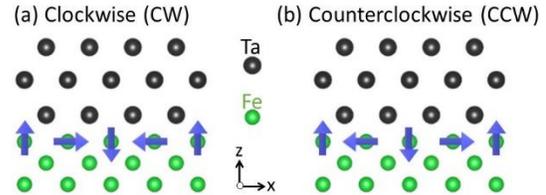

*Fig. 2 Fe/Ta supercell structures with spin spirals rotating in the $(xz)$-plane used to simulate (a) CW and (b) CCW configurations. The DMI energy is calculated as the difference in the total energy for those configurations. For clarity, the spins are shown only for one Fe layer.*

## III. RESULTS
### A. Experimental results

The studied system is a sputtered Ta/FeCoB/TaOx trilayer with weak DMI strength (< 300 µJ/m²). A homogeneous bottom Ta/FeCoB interface is achieved by depositing the Ta layer at constant thickness, while crossed wedges of FeCoB and top Ta layers create gradients: FeCoB thickness varies along the x-axis and the oxidation state of TaOx along the y-axis, since the top Ta is oxidized after deposition (Fig. 1(b)). Hence, for a given ferromagnet thickness, we obtain different oxidation states at the top FeCoB/TaOx interface. Conversely, for a fixed oxidation state, the ferromagnet thickness varies continuously.

We first discuss the effect of ferromagnetic thickness ($t$) and the degree of oxidation at the FeCoB/TaOx interface on the magnetic anisotropy. The effective anisotropy, $K_{eff} = \frac{K_s}{t} - K_d$, depends on the surface anisotropy constant $K_s$ and the dipolar $K_d = \frac{1}{2}\mu_0 M_s^2$ anisotropy constants. Variations in $t$ directly influence $K_{eff}$, while changes in oxidation state impact interfacial properties ($K_s$).

At a critical ferromagnet thickness $t_c$, $K_d$ overcomes the surface contribution $\frac{K_s}{t}$, leading to a sign change of $K_{eff}$ which corresponds to the crossover between PMA and In-Plane (IP) anisotropy. For ultrathin ferromagnet, size effects and dead layers induce a paramagnetic (PM) regime at room temperature [20,21].

*Contact author: johanna.fischer@cea.fr

†Contact author: helene.bea@cea.fr

Magnetometry measurements using the polar magneto-optic Kerr effect (p-MOKE) reveal square hysteresis loops in PMA regions (dark grey in Fig. 3(a)). Remanence loss is obtained in the surrounding intermediate grey regions arising from labyrinthine or stripe domain formation at zero magnetic field due to reduced $K_{eff}$. As t increases, the system transitions into the IP anisotropy region (light gray in Fig. 3(a)), where $K_{eff}$ changes sign. The proximity of remanence loss and $K_{eff}$ transition regions highlights remanence as a reliable indicator of anisotropy changes. Now starting from PMA region and decreasing $t$ leads to a PM region with $K_{eff}$ and saturation magnetization dropping to zero, also inducing remanence loss.

Both wedges in the system induce anisotropy transitions, with the PMA-to-IP transition occurring at FeCoB thickness of approximately 1.1–1.2 nm. Additionally, for top Ta thinner than 0.9 nm, the PMA region narrows with decreasing Ta thickness. This indicates partial oxidation of FeCoB, creating a FeCoB/FeCoBOx/TaOx top interface. On the other hand, for Ta > 1 nm, increasing Ta thickness reduces the PMA region, leading to a PM regime. The resulting underoxidized FeCoB/Ta/TaOx interface causes a loss of magnetism induced by Fe/Ta orbital hybridization [30]: our *ab initio* calculations find a 20% reduction of the magnetic moment of the Fe atom at the Ta interface (Supplementary 2). The interfacial disorder in experimental samples may also reduce ferromagnetic order.

In the low remanence regions showing demagnetized state at zero magnetic field, the application of a small perpendicular magnetic field (< 1 mT) allows to nucleate skyrmions. Brillouin Light Scattering (BLS) measurements demonstrated a DMI sign inversion with the oxidation of the FeCoB/TaOx interface (on both sides of the red star in Fig. 3(a)) [19], resulting in the emergence of skyrmions with opposite chirality on each side of this transition.

To better understand the DMI behavior and its dependence on the FeCoB/TaOx interface oxidation state and FeCoB layer thickness, we map the chirality of the spin textures in the PMA region. The sign of DMI coefficient in a large number of wafer positions is measured from the direction of the current-induced motion of DWs, observed with polar MOKE microscopy: CW (resp. CCW) DW moves along (resp. opposite to) the current direction when using a HM with a negative spin Hall angle like Ta [21]. This systematic study of the influence of FeCoB thickness and FeCoB/TaOx oxidation state on DMI chirality over the whole PMA region is presented in Fig. 3(a), with examples of MOKE images in Fig. 3(b). Each square in Fig. 3(a) indicates a position where DW motion under current was analyzed to determine the chirality and the DMI sign. The regions exhibiting CCW and CW chirality are depicted in orange and purple, respectively. The red zero DMI line indicates the DMI chirality transition, at which nearly no domain wall motion is visible. This line joins the red cross and star previously observed in the skyrmion region [19]. For a fixed FeCoB thickness (vertical line in Fig. 3(a)), DMI chirality changes with top Ta thickness and oxidation state, consistent with [16,18]. Surprisingly, we observe a DMI chirality inversion for a fixed top Ta thickness when varying FeCoB thickness (horizontal line in Fig. 3(a)). This unexpected dependence of DMI chirality on FeCoB thickness is evident across several FeCoB/TaOx oxidation states.

The effective DMI in our sample is considered as the sum of contributions from both interfaces [31,32]. The sign of the DMI contribution from the top FM/oxide interface was shown to be dependent on oxidation [16–18]. The DMI contribution from the bottom Ta/FeCoB interface is around $D = 20 - 30\ \mu J/m^2$ [19] which corresponds to $D_s = 20 - 30\ fJ/m$ [33,34]. BLS measurements in low remanence regions revealed DMI values around $D = -100\ \mu J/m^2$ near the PM/PMA transition (slightly thinner top Ta than the red cross in Fig. 3(a)) [35] and $+ 100\ \mu J/m^2$ ($- 100\ \mu J/m^2$, respectively) near the PMA/IP transition for top Ta thicker (thinner resp.) than the red star [19,20], close to the limit of resolution of the technique. For these three measurements, removing the bottom interface contribution showed that the top FeCoB/TaOx DMI contribution changes sign [19]. Hence, the sign of the surface DMI, $D_s$, originating from the top FM/MOx interface, depends on the oxidation degree. Remarkably, $D_s$ sign also changes with the FM thickness, a result that has not yet been unambiguously reported experimentally.

*Contact author: johanna.fischer@cea.fr

†Contact author: helene.bea@cea.fr

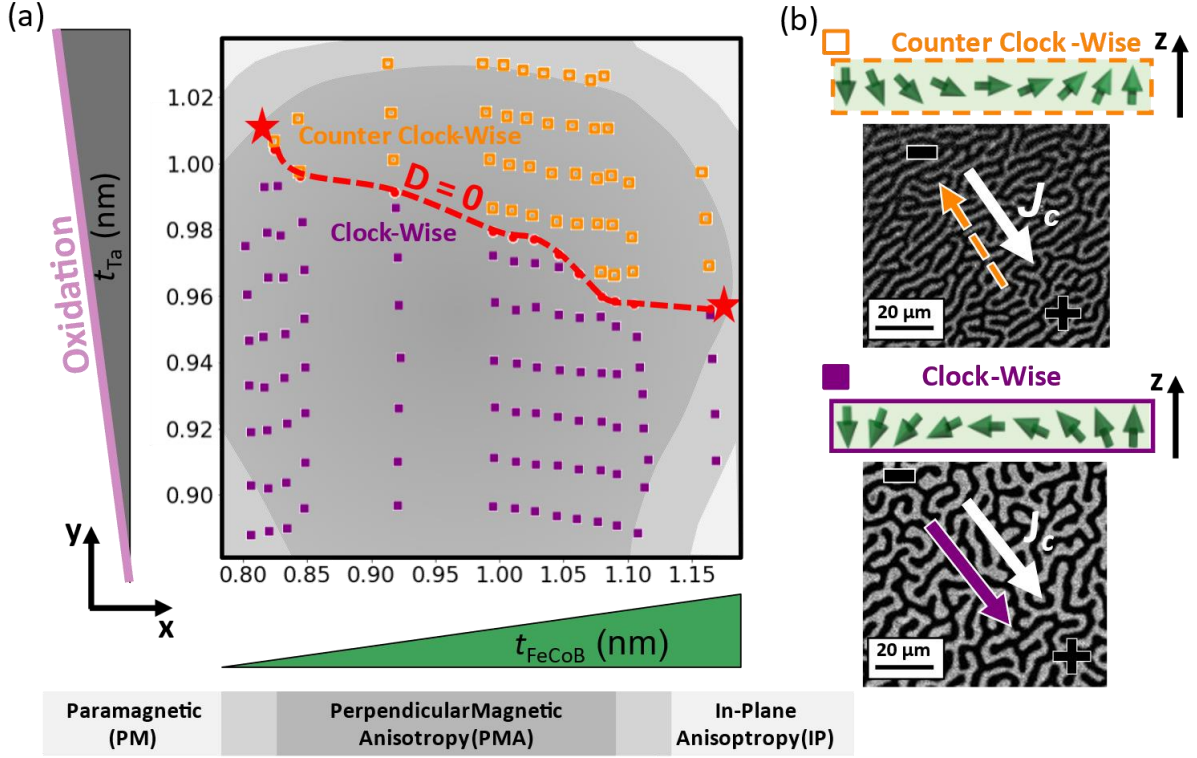

*Fig. 3 (a) Reconstructed experimental map of the DW chirality in the PMA region (dark grey) with respect to FeCoB thickness ($t_{FeCoB}$) and oxidation state ($t_{Ta}$). At the border of PMA region (intermediate grey), spontaneous demagnetization occurs at zero magnetic field, creating domains and DWs. In the PMA region DWs can be created by a proper magnetic field sequence, but they are much less dense. Measured points with CW (resp. CCW) DWs are represented with full purple squares (resp. empty orange squares). Red dashed line represents DMI sign crossover and corresponds to no DW motion. It meets the DMI sign inversion observed in skyrmion regions (red star and cross) [19]. (b) Differential MOKE images for both domain wall chiralities. The current direction $J_c$ and the DW motion are represented by a white and orange/purple arrow respectively.*

## B. *Ab initio* calculations

To understand the observed DMI sign inversion with the FM thickness, we conducted *ab initio* calculations. $Fe_n/Ta_m$ bilayer with variable Fe thickness and Ta oxidation state, n and m representing the number of monolayers MLs (Fig. 1(a)), is used to model the top interface of our sample, FeCoB/TaOx. A justification of the use of the bilayer only and not the full trilayer for the calculation is provided in Supplementary 3. The calculated DMI energy $E_{DMI}$ is proportional to the interfacial DMI coefficient $D_s$. We have explored two different scenarios to model the oxidized Ta wedge in the experimental samples.

In the first scenario, only the interfacial Ta layer was oxidized and the oxidation percentage was varied (0-100%) (Fig. 4(b-d)) for different Fe thickness (3-9 MLs). The calculated DMI energies are presented versus Fe thickness and Ta oxidation percentage in Supplementary 4. Fig. 4(a) summarizes the $E_{DMI}$ where different effects can be distinguished. First, $E_{DMI}$ is highly dependent on the Ta oxidation percentage for all Fe thicknesses. Notably, for 5 to 9 Fe MLs, the DMI changes sign at low oxidation percentages below 25% beyond which its value increases. This DMI sign change with Ta oxidation is qualitatively consistent with the experimental map in *Fig. 3*. *Ab initio* calculations deal with single crystalline structures, whereas our experimental samples are polycrystalline, which might explain an offset in transition thickness values.

Importantly, a DMI sign change between three and five Fe monolayers is found for the unoxidized interface. This is qualitatively consistent with our experimental observation of DMI sign change for increased FM layer thickness at small and constant

*Contact author: johanna.fischer@cea.fr

†Contact author: helene.bea@cea.fr

FeCoB/TaOx oxidation state (upper part of the map in *Fig. 3*(a)). For higher percentages of Ta oxidation, no DMI sign change with Fe thickness is found. Finally, the simultaneous oxidation of interfacial Ta and Fe (Fig. 4(d)) tends to decrease the DMI values and a further decrease occurs for thicker Fe layers. This shall correspond to thinner Ta regions than studied in Fig. 3. The small DMI predicted in this overoxidized region is interesting for a potential reversal of the chirality by applying a gate voltage [20], but is beyond the scope of the present study.

In the second scenario, the oxygen position with respect to the Fe/Ta interface can be an additional model (besides the oxygen percentage at the interface) to describe the oxidized Ta wedge in the experimental samples. In Fig. 4(e), the calculated $E_{DMI}$ values as a function of Fe thickness are shown for different oxygen distances from the Fe interface ranging from non-oxidized to fully oxidized Ta layers (Fig. 4(f-i)). Here, a five monolayer thick Ta was chosen so that to have more data points along the path of oxygen towards the Fe interface. Interestingly, the DMI sign reversal as a function of Fe thickness is now present for several oxygen positions.

Again, a quantitative comparison with experiments remains difficult since the experimental effective DMI includes the small contribution from the bottom Ta/FeCoB, independent on the FeCoB or Ta top thicknesses [19]. Despite that matching the thickness region for DMI reversal in experiment and calculations is not exact, this does not affect the underlying mechanism described hereafter (Supplementary 5).

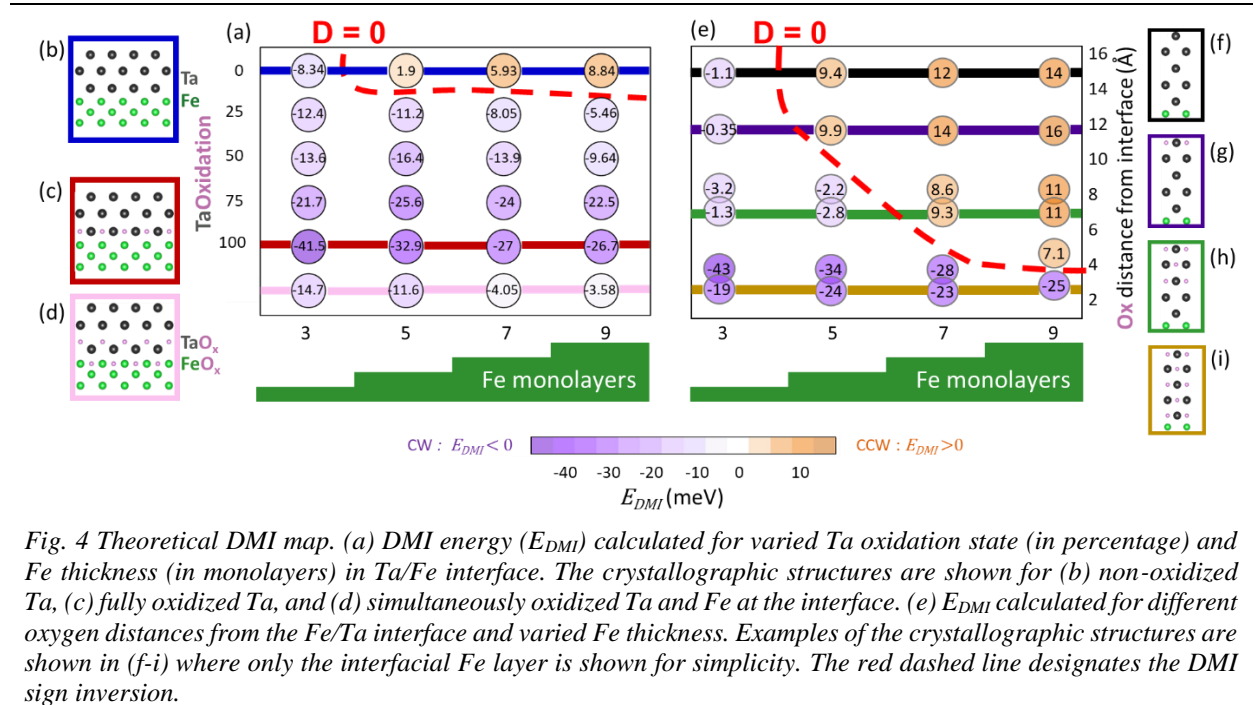

*Fig. 4 Theoretical DMI map. (a) DMI energy ($E_{DMI}$) calculated for varied Ta oxidation state (in percentage) and Fe thickness (in monolayers) in Ta/Fe interface. The crystallographic structures are shown for (b) non-oxidized Ta, (c) fully oxidized Ta, and (d) simultaneously oxidized Ta and Fe at the interface. (e) $E_{DMI}$ calculated for different oxygen distances from the Fe/Ta interface and varied Fe thickness. Examples of the crystallographic structures are shown in (f-i) where only the interfacial Fe layer is shown for simplicity. The red dashed line designates the DMI sign inversion.*

As the oxidation effect on the DMI has been studied before in similar systems [16–18,36], we focus here on the mechanism underlying the dependence of the DMI on the FM thickness at the Fe/Ta interface. Due to the interfacial character of $D_s$, it is unexpected that its sign may depend on the FM thickness. Fig. 5(a) presents the layer-resolved difference in the chirality-dependent spin-orbit coupling (SOC) energy i.e., $\Delta E_{soc}^k = E_{cw}^k - E_{ccw}^k$ where k is the atomic layer. As expected from the Fert-Levy DMI mechanism [8], the SOC energy is predominantly located at the interfacial Ta1 layer. Interestingly, it shows a remarkable sign crossover between $Fe_3Ta_3$ and $Fe_5Ta_3$ structures, which is at the origin of the overall DMI sign change. Noteworthy, the Ta2 and Ta3 layers hold minor and cancelling contributions to the overall $\Delta E_{soc}$ which justifies the choice of a three-monolayer thick Ta layer in our calculations (Supplementary 6).

*Contact author: johanna.fischer@cea.fr

†Contact author: helene.bea@cea.fr

Of note, the Fe/vacuum interface has a non-negligible contribution to $E_{DMI}$. For $Fe_5Ta_3$, $Fe_7Ta_3$ and $Fe_9Ta_3$ in Fig. 5(a), the three Fe atoms close to vacuum have similar $\Delta E_{soc}$. However, for the ultra-thin $Fe_3Ta_3$, the vacuum contribution cannot be disentangled from the Fe/Ta interface contribution, which makes this correction difficult to perform. Overall, the vacuum interface contribution to the total $E_{DMI}$ might shift a bit the Fe thickness at which the DMI changes sign. Nevertheless, the proposed mechanism of the inversion of the $\Delta E_{soc}$ of Ta1 mainly contributing to the DMI sign inversion, remains valid.

A microscopic insight into the DMI dependence on the FM thickness is acquired by comparing the variation of 5$d$ orbital-resolved SOC energy of interfacial Ta1 layer in Fe$_3$Ta$_3$ and Fe$_5$Ta$_3$ (Fig. 5(b,c)). A substantial difference is observed in the SOC matrix element contribution corresponding to the hybridization between $d_{z^2}$ and $d_{xz}$ orbitals, strongly negative for Fe$_3$Ta$_3$ and positive for increasing Fe thickness.

*Contact author: johanna.fischer@cea.fr

†Contact author: helene.bea@cea.fr

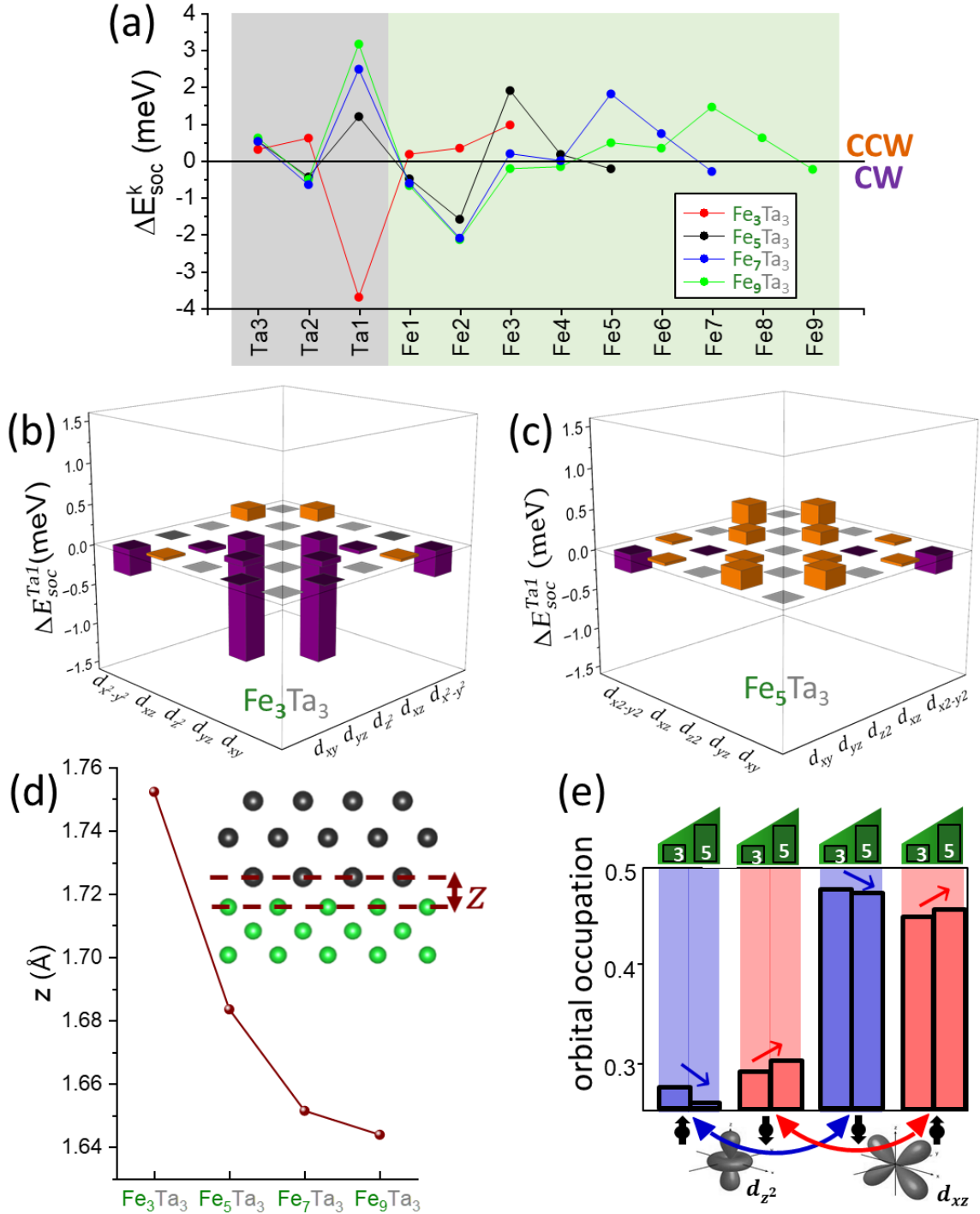

Fig. 5. Mechanism of the DMI dependence on the FM thickness. (a) Layer resolved difference in the chirality-dependent spin-orbit coupling energy $\Delta E_{soc}^{k}$ calculated for Ta/Fe structures with variable Fe thickness. Variation of 5d orbital-resolved SOC energy of interfacial Ta1 layer in (b) $Fe_3Ta_3$ and (c) $Fe_5Ta_3$. (d) Variation of the interfacial distance 'z' between Ta1-Fe1(inset) for the different calculated structures. (e) The spin dependent d-orbital occupations ($d_{z^2}$ and $d_{xz}$) of Ta1 in $Fe_3Ta_3$ and $Fe_5Ta_3$, referred to as 3 and 5.

*Contact author: johanna.fischer@cea.fr

†Contact author: helene.bea@cea.fr

In the framework of the first-order perturbation theory [37,38], the occupied $5d$ orbital states of the interfacial Ta1 layer mainly contribute to the DMI. Consequently, the corrections to the total energy due to DMI can be approximated by the expectation value of the SOC operator as $\langle \psi_{lm,s} | \hat{H}_{soc} | \psi_{lm,s} \rangle$ where $|\psi_{lm,s}\rangle$ represents the spin-dependent occupied states of Ta1. Considering the spin-mixing transition terms between fully-occupied $d$-orbitals, the highest contribution to DMI is given by the expectation value of the SOC matrix elements $\langle d_{z^2}^\uparrow + d_{xz}^\downarrow | \hat{H}_{soc} | d_{z^2}^\uparrow + d_{xz}^\downarrow \rangle = -\sqrt{6}$ and $\langle d_{z^2}^\downarrow + d_{xz}^\uparrow | \hat{H}_{soc} | d_{z^2}^\downarrow + d_{xz}^\uparrow \rangle = \sqrt{6}$ [38]. This is consistent with our finding from orbital-resolved SOC energies where $d_{z^2}^\uparrow + d_{xz}^\downarrow$ contributes to the largest change (Fig. 5(b,c)). However, to explain the sign change it is relevant to compare the d-orbital occupations of Ta1 in Fe$_3$Ta$_3$ and Fe$_5$Ta$_3$ shown in Fig. 5(e). On one hand, the decrease of both the $d_{z^2}^\uparrow$ and $d_{xz}^\downarrow$ occupation leads to a decrease in the $\langle d_{z^2}^\uparrow + d_{xz}^\downarrow | \hat{H}_{soc} | d_{z^2}^\uparrow + d_{xz}^\downarrow \rangle$ matrix element when increasing the Fe thickness from three to five monolayers. At the same time, the increase of the $d_{z^2}^\downarrow$ and $d_{xz}^\uparrow$ occupation yields an increase in the $\langle d_{z^2}^\downarrow + d_{xz}^\uparrow | \hat{H}_{soc} | d_{z^2}^\downarrow + d_{xz}^\uparrow \rangle$ matrix element. Simultaneously, these two effects add a positive contribution to the $\Delta E_{soc}$ of the interfacial Ta1 layer, which in turn dictates the overall CW to CCW DMI sign crossover with the increase in Fe thickness (detailed calculation of the interlayer and intralayer contributions to DMI in supplementary 7).

The change in the $d$-orbital occupation can be attributed to the structural relaxation upon increasing the FM thickness. In Fig. 5(d), as the Fe thickness increases, the interfacial distance $z$ between Ta1-Fe1 decreases. It triggers the change in the occupation of the overlapping $d$-orbitals, mainly the out-of-plane $d_{z^2}$ and $d_{xz}$. This affects the DMI energy, as seen above, but also the PMA and the magnetic moment (Supplementary 2). We note that the variation of the interfacial distance $z$ is led by the Fe-Fe interlayer distances that vary notably when increasing the Fe thickness (Supplementary 8).

To further support the fact that structural relaxation can induce a DMI sign reversal independent of the oxidation state of Ta at Fe/Ta interface, we show in Fig. 6 the variation of the interfacial distance z between Ta1-Fe1 as a function of increasing Fe thickness for different oxidation states (same as in Fig. 4(e)). It can be seen that z decreases with increasing the Fe thickness for all oxidation states. To correlate this structural change with the DMI sign reported in Fig. 4(e), we added a dashed red line corresponding to the DMI sign reversal. We see that its position is consistent with the DMI sign transition found between Ta$_3$Fe$_3$ and Ta$_3$Fe$_5$ for z around 1.71Å in Fig. 5(d). In fact, we can distinguish between two regimes. In the first, the oxygen is far away from the interface (Fig. 6(a)), thus the only mechanism decisive to the DMI sign reversal is the orbital filling variation induced by structural relaxation as discussed earlier. Consequenlty, we observe a good correspondence between the line sketched at a certain z value and the DMI sign calculated for different oxidation states. However, in the regime were the oxygen starts to be relatively close to the interface (Fig. 6(b)), an additional mechanism comes into play which is the Fe-O orbital hybridization. This is well-known to strongly affect the DMI sign. In this case, the interfacial distance z does not solely determine the DMI sign and we thus observe that the dashed line fails to correlate with the DMI sign namely in Fe$_5$Ta$_5$ and Fe$_7$Ta$_5$ where O is placed at Ta5, Ta4, Ta3, Ta2.

Noteworthy, the DMI values calculated as a function of Fe thickness in both scenarios of TaOx i.e., oxidizing Ta1 layer only (red solid line in Fig. 4(a)) or all the Ta layers (yellow solid line in Fig. 4(e)), are in good correspondance which highlights the interfacial aspect of DMI.

It is important to point out that in the ultrathin ferromagnet thickness (Fe$_3$Ta$_3$) as discussed from the structural parameters and layered resolved SOC energies, an additive scheme of DMI of the two interfaces cannot be applied since the two interfaces cannot be fully decoupled. Therefore, we do not exclude that, in the ultrathin limit, the coupling of the bottom and top interfaces might be an additional or alternative mechanism at the origin of the DMI reversal measured in the samples beside the structural relaxation effect described earlier.

We stress that this DMI sign change with FM thickness originates from spin-orbit coupling in the unoxidized or underoxidized top Ta heavy metal and is thus attributed to a Fert-Levy DMI mechanism. Recently a Rashba-type DMI sign change was predicted at the Fe/MgO interface with increasing Fe thicknesses from 1 to 5 monolayers [39]. The change of DMI sign with FM thickness in the ultrathin regime might thus occur in other systems, which would be interesting to explore experimentally.

*Contact author: johanna.fischer@cea.fr

†Contact author: helene.bea@cea.fr

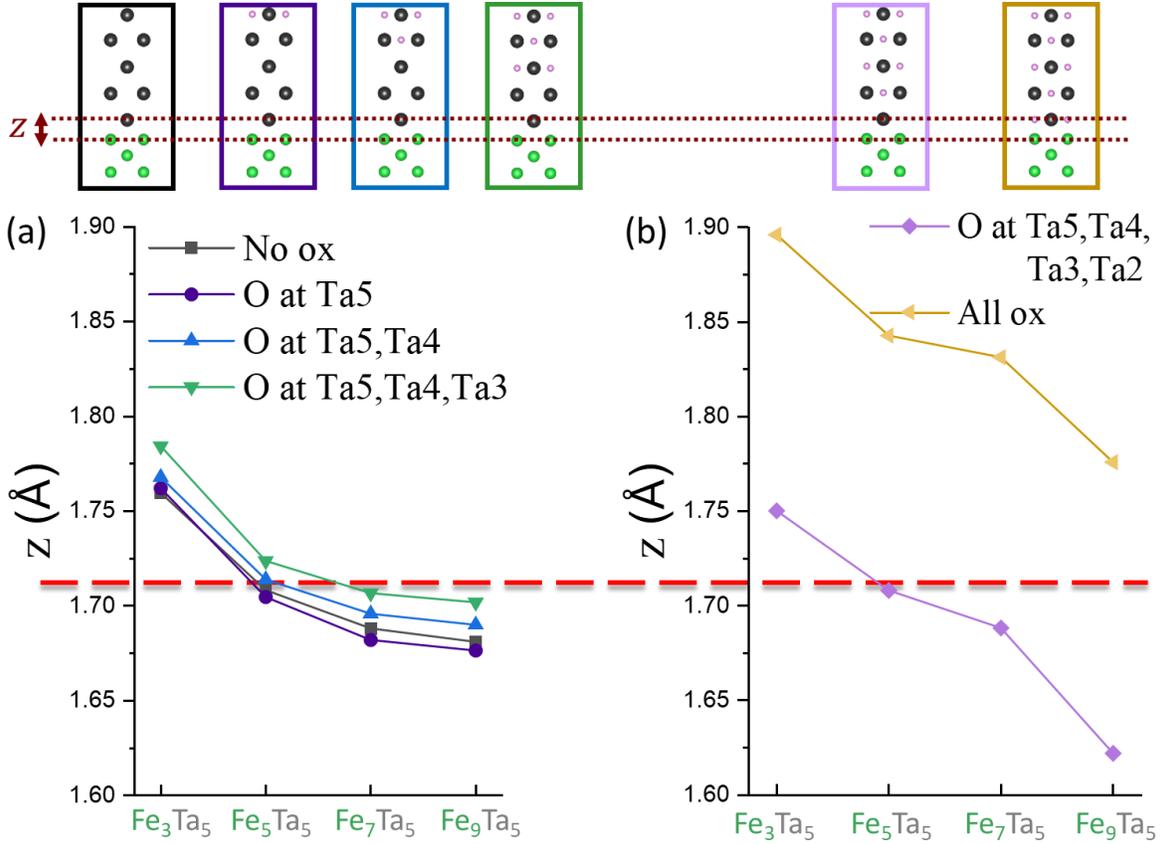

*Fig. 6. Structural relaxation with varied Fe thickness and oxidation state. The top crystal structures show the models used to describe the gradual oxidation of the Ta-Fe interface. z designates the interfacial Fe-Ta distance. The variation of z for different oxygen positions from the interface as a function of increased Fe thickness is shown in two splitted regimes where oxygen is relatively (a) away and (b) close from the interface. The sketched dashed red line corresponds to the z value at which the DMI is supposed to change sign in correlation with Fig. 4(e).*

## IV. CONCLUSIONS

To summarize, we have conducted a combined experimental and *ab initio* study in a heavy metal/ferromagnet/oxide trilayer with weak DMI. We have analyzed separately the influence of the oxidation state at the top FM/MOx interface and the FM thickness on the DMI chirality. Here, we unveil unprecedented dependence of the interfacial DMI chirality on the thickness of the FM layer for an under-oxidized FM/M(Ox). Our *ab initio* calculations corroborate our experimental findings and propose the orbital filling variation induced by structural relaxation as the underlying mechanism. This is directly linked to a strain effect, thereby unlocking exciting opportunities for manipulating chiral domain walls and skyrmions through strain engineering. Timely, a novel research area is emerging where skyrmions may be nucleated and propagated using transient strain provided by surface acoustic waves [40–42].

## ACKNOWLEDGMENTS

The authors acknowledge funding by the French ANR (contracts ADMIS n°ANR-19-CE24-0019, CHIREX n°ANR-22-EXSP-0002, SPINTHEORY n°ANR-22-EXSP-0009). This project has received funding from the European Union's Horizon 2020 research and innovation program under Grant Agreement No. 800945 (NUMERICS–H2020-MSCA-COFUND-2017).

*Contact author: johanna.fischer@cea.fr

†Contact author: helene.bea@cea.fr

*Contact author: johanna.fischer@cea.fr

†Contact author: helene.bea@cea.fr

*Contact author: johanna.fischer@cea.fr

†Contact author: helene.bea@cea.fr

# Dzyaloshinskii-Moriya interaction chirality reversal with ferromagnetic thickness

Capucine Gueneau[1,¶], Fatima Ibrahim[1,¶], Johanna Fischer[1,*], Libor Vojáček[1], Charles-Élie Fillion[1], Stefania Pizzini[2], Laurent Ranno[2], Isabelle Joumard[1], Stéphane Auffret[1], Jérôme Faure-Vincent[1], Claire Baraduc[1], Mairbek Chshiev[1,3], and Hélène Béa[1,3*]

[1]Univ. Grenoble Alpes, CEA, CNRS, Spintec, 38000 Grenoble, France
[2]Univ. Grenoble Alpes, CNRS, Néel Institute, 38042 Grenoble, France
[3]Institut Universitaire de France (IUF), 75000 Paris, France

¶ C.G. and F.I. contributed equally
* Corresponding authors, Johanna.fischer@cea.fr, Helene.bea@cea.fr

**Supplementary Material:**

1. <u>Reconstruction of the thickness maps due to wedge deposition</u>

The deposition technique is schematized in the following Figure S1. The wedges are obtained by off-centering the sample with respect to the target as represented in the Figure S1(a). It leads to a relatively homogeneous thickness when the substrate is facing the target, while a linear variation of thickness is obtained for off-centered substrate: the thickness, represented with the black line in Figure S1(a), has been characterized by X ray reflectometry (XRR).

The target consists of a disk made of the deposited material. As a result, the deposition thickness forms concentric circles of constant thickness, as depicted in the top view in Figure S1(b).

To construct the DMI sign map as a function of thickness, it is necessary to convert the coordinates on the wafer into corresponding thickness values. In other words, we must transform the circular thickness distribution, represented in Figure S1(c), into a flattened thickness map. To achieve this, we assume that the deposited thickness $t$ varies linearly with the distance from the target to the center (red curve in Figure S2(a)).

However, in reality, the deposition follows a third-degree polynomial distribution (green curve in Figure S2, experimentally measured by X-ray reflectometry). The sample is centered at position 100 mm, with the thicker deposition region located at 50 mm and the thinner region at 150 mm. To evaluate the impact of the linear approximation on thickness estimation, we compute the normalized difference between the two curves. At the sample edges, this approximation introduces at maximum a 5% error, while in the region of interest where the DMI changes sign with the FM thickness, positioned around 80, the error is reduced to just 1%.

Between the two points marking the DMI sign change in Figure 3 of the main text, the difference in Ta top-layer normalized thickness is 5%. Importantly, the 1% error in Ta thickness determination remains smaller than the observed variation along the DMI sign change line. This confirms that there is no significant change in the oxidation state at the interface along the FeCoB wedge direction (ie. corrected according to the circular deposition).

Also, the expected alignment of the target with respect to the sample is of 1°, this error will not introduce a significant change in Ta thickness along the FeCoB wedge direction (again, if we correct with the deposition with circular iso-thickness lines).

Finally, this change in sign with the ferromagnet thickness has also been seen in other studies in our team [4] for samples with slightly different thicknesses. It changed slightly the position of the whole PMA region on the wafer, but this change of DMI sign with ferromagnetic thickness was also found. This is thus reproducible and not very sensitive to the exact position on the wafer, ie not an effect of the edge of the wafer: it is thus really related to the change of ferromagnetic thickness.

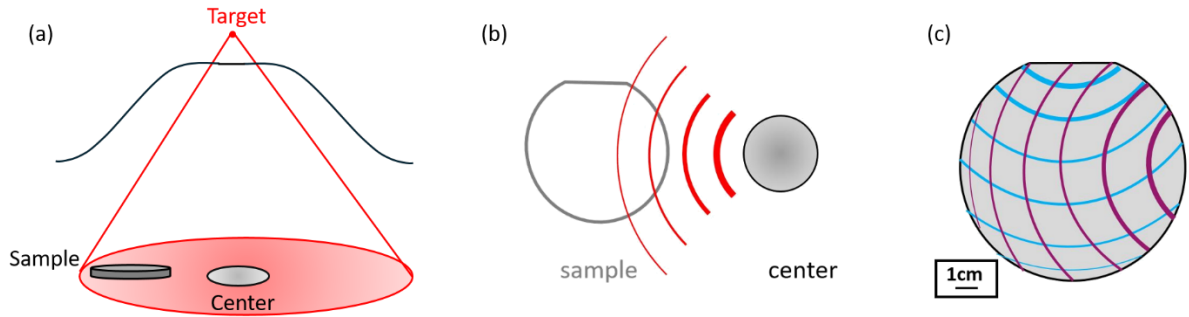

Figure S1. (a) Side view schematics of the target-substrate deposition geometry. The black line gives the variation of deposited thickness in the radial direction. (b) top view of the target/substrate geometry. (c) reconstruction of the iso-thicknesses along circles on the wafer for crossed wedges (purple and blue circular arcs for the 2 materials, thicker lines correspond to thicker material)

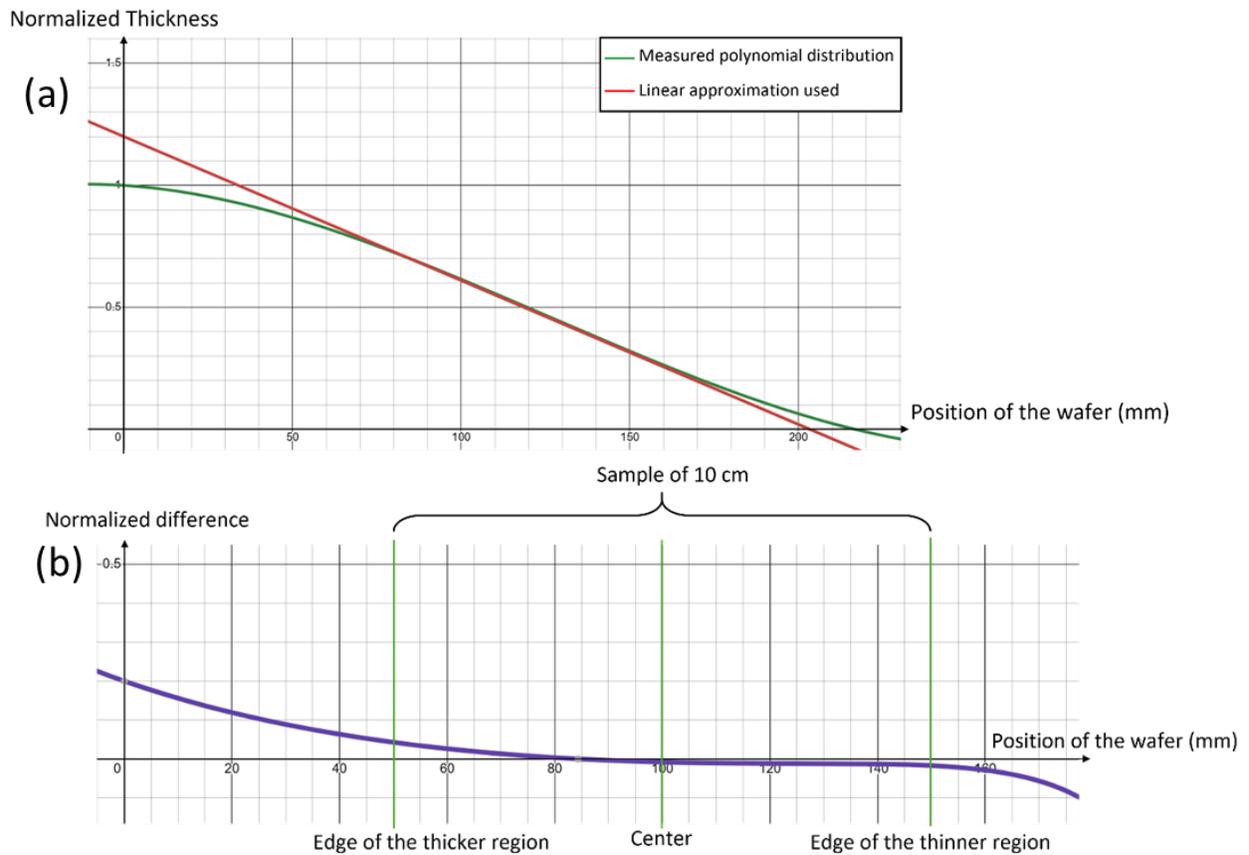

Figure S2. (a) Variation of thickness deposited on an off-centered wafer. Position 0 corresponds to the center. Red line is the fit used for reconstructing remanence maps while green curve is a polynomial fit corresponding to the XRR characterized thickness variation measured between 50 and 150mm. (b) Normalized difference of thickness between the approximation and the exact distribution depending on the position with respect to the target center. During off-axis deposition, the wafer lies between position 50 and 150mm. The maximum error in our reconstructions is below 5%

2. <u>Magnetic moments in Ta/Fe interface</u>

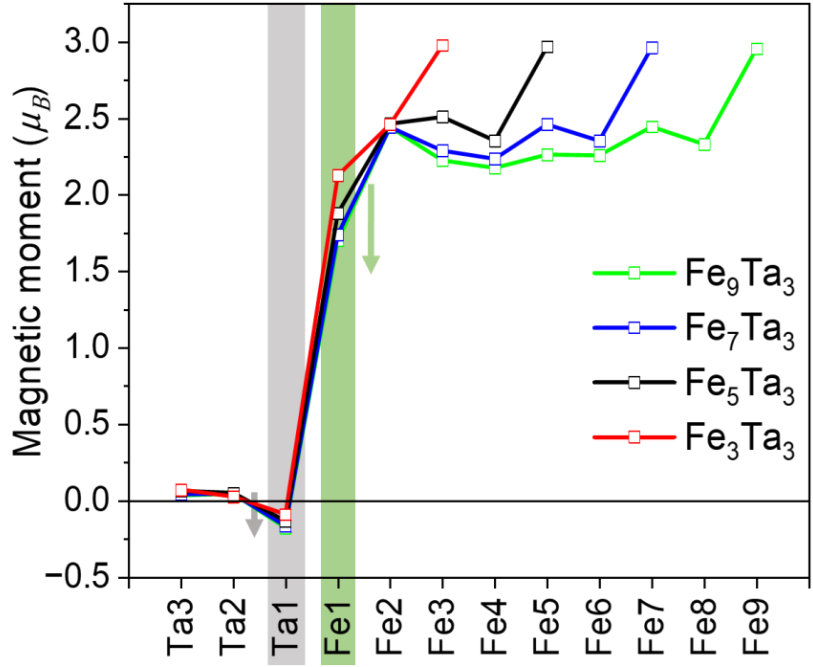

*Figure S3 The magnetic moment per atom calculated for different Fe thicknesses in Fe/Ta interface. The green (gray) arrow represents the reduction (increase) of the Fe1 (Ta1) moment with the increase of Fe thickness.*

As discussed in the main text, the structural relaxation induced by increasing the Fe thickness has a direct impact on the magnetic properties of Fe/Ta interface.

Figure *S3* shows the magnetic moment per atom calculated for different Fe thicknesses. The interfacial Fe1 atom has a reduced magnetic moment as the Fe layer gets thicker; a reduction of about 20% is found between $Fe_3Ta_3$ and $Fe_9Ta_3$ structures. On the other hand, the magnetic moment of the interfacial Ta1 atom increases negatively. Those two behaviors are mediated by the change in the *d*-orbital occupations, which in turn orginates from the decrease of the interfacial distance '*z*' as shown in Figure 5(d) of the main text. For instance, Figure S4 shows the projected density of states (PDOS) of the $d_{z^2}$ and $d_{xz}$ orbitals of the Ta1 atom compared for $Fe_3Ta_3$ and $Fe_5Ta_3$. It can bee seen that the splitting between the occuppied majority and minority spin states increases leading to the increase of the overall magnetic moment of Ta1 with the Fe thickness. The occupied states below the Fermi energy ($E_F$) are integrated to obtain the orbital occupation presented in the main text and used in the framework of the first-order perturbation theory analysis.

It is also important to point out that the Fe/Ta interface features a decrease of the interfacial Fe1 atom (observed for all Fe thicknesses considered) compared to its bulk value (2.2 $\mu_B$). This is consistent with the experimental observation that FeCoB becomes paramagnetic when too thick unoxidized Ta is deposited atop (see Figure S3 and [1]). At the same time, a small negative magnetic moment is induced on the interfacial Ta1 atom.

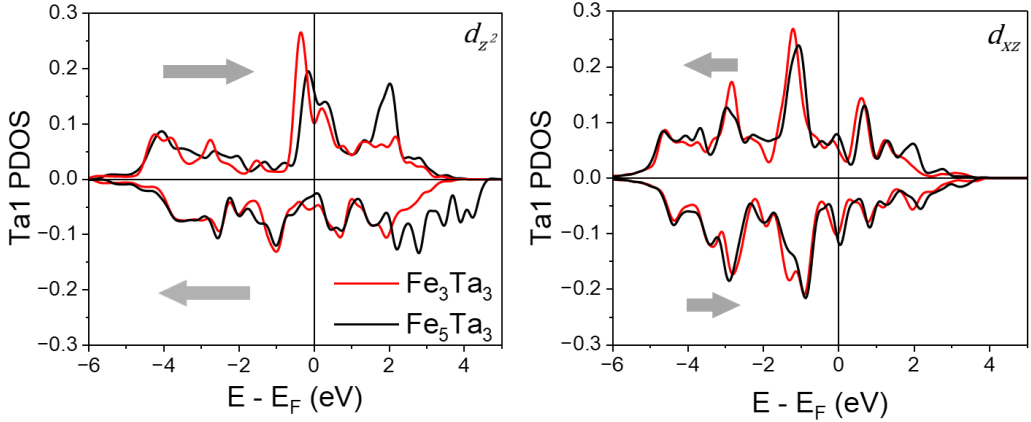

*Figure S4 Projected density of states (PDOS) of the $d_{z^2}$ and $d_{xz}$ orbitals of the interfacial Ta1 atom compared for Fe$_3$Ta$_3$ and Ta$_3$Fe$_5$. The occupied states below the Fermi energy ($E_F$) are integrated to obtain the orbital occupation presented in the main text. The weights of the arrows designate the size of the shift of the occupied states when increasing the Fe thickness from Fe$_3$Ta$_3$ to Fe$_5$Ta$_3$. Positive (negative) Ta1 PDOS correspond to majority (minority) spins, respectively.*

3. <u>Discussion on the DMI in trilayers</u>

As explained in the manuscript, the two experimental interfaces are not equivalent due to the order of the sputtering deposition: due to its larger mass, the Ta sputtered onto the FeCoB leads to a more intermixed interface that than the FeCoB onto the bottom Ta layer. Similarly, Pt/Co/Pt nominally symmetric trilayers have been shown to display non zero experimental DMI values (for instance in [5]). Thus, calculating Ta/Fe/Ta trilayer would not be relevant since symmetry leads to a vanishing total DMI.

We have shown previously that in this Ta/FeCoB/TaOx system the effective DMI is mainly coming from the top FeCoB/TaOx interface and that the bottom interface contribution has a smaller amplitude [4]. Besides, the intermixed layer in the top interface shall not vary with the FeCoB thickness which further supports that the mechanism behind the DMI sign reversal is the structural relaxations mediated by the variation in the orbital fillings.

Moreover, in the ultrathin ferromagnet thickness (Fe$_3$Ta$_3$) as discussed from the structural parameters and layered resolved SOC energies, an additive scheme of DMI of the two interfaces cannot be applied since the two interfaces cannot be decoupled. Therefore, we do not exclude that the coupling of the bottom and top interfaces might be an additional mechanism at the origin of the DMI reversal measured in the samples beside the structural relaxation effect described earlier.

4. <u>DMI variation with Ta oxidation and Fe thickness in Fe/Ta interface</u>

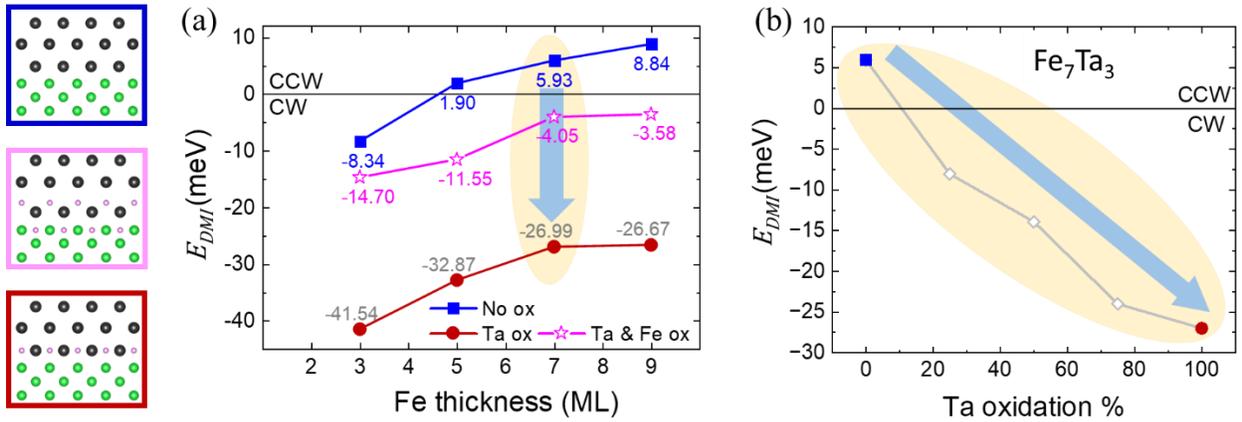

*Figure S5 (Left)The crystallographic structures used for the three extreme cases considered for the DMI calculation are shown for non-oxidized Ta, fully oxidized Ta, and simultaneously oxidized Ta and Fe interface. (a) The DMI energy variation with the Fe thickness for different oxidation states of Fe/Ta interface. (b) The DMI energy variation as a function of the Ta oxidation percentage for a particular Fe thickness of seven monolayers. These data are summarized in Figure 4(a) of the main text.*

Here, we present the detailed results of the DMI calculations. Figure S5(a) shows the variation of the DMI energy as function of the Fe thickness in $Fe_nTa_3$ interface calculated for different Ta oxidation states. First in the absence of oxygen, the square blue line shows that the DMI changes sign between three and five Fe monolayers. In the second extreme case where the Ta interfacial layer is fully oxidized, the DMI preserves its negative sign but decreases in value for thicker Fe. The case of overoxidation is modelled by fully oxidizing both the interfacial Fe and Ta layers (magenta stars in Figure S5(a)). Interestingly, the oxidation of the Fe layer decreases the DMI values compared to the case where only Ta is oxidizied. To demonstrate the DMI dependence on the Ta oxidation, we plot in Figure S5(b) the variation of the DMI energy as function of the oxygen percentage of the Ta interfacial layer for the particular seven monolayer Fe thick system $Fe_7Ta_3$. The DMI changes sign at low oxidation percentages (below 25%) beyond which its value increases with the increase of the oxygen percentage. The full map of the DMI energy varying with both Ta oxidation percentage and the Fe thickness is presented in Figure 4(a) of the main text.

5. Thickness ranges in experiment versus *ab initio*

The change of DMI sign in our *ab initio* calculations is obtained between 3 and 5 monolayers of Fe for no oxidation, which corresponds to thickness variation between 0.28 and 0.56 nm or between 5 and 7 (7 and 9 respectively) when oxygen lies in the third or fourth Ta layer (second Ta layer, resp.). If we consider a typical dead layer thickness of about 0.25 to 0.6nm [3,4], which is included in the nominal experimental thickness, both are thus in a similar range. The *ab initio* calculations are not expected to exactly fit the experimental results since the systems are not fully identical: experimental samples are polycrystalline whereas structures are epitaxial in *ab initio*, target experimental composition is Fe-rich Fe72Co8B20 modeled by pure Fe layers in *ab initio*. However, the goal of the *ab initio* calculations is to provide a possible explanation for the mechanism behind the change of DMI sign with the ferromagnetic thickness variations. Despite that matching between the thickness ranges in experiment and calculations is not exact, the Fe thickness at which the DMI changes sign might shift when considering the vacuum correction (as discussed in the main text). However, this will not affect the mechanism i.e., the variations in orbital filling and inter-atomic distances at the interface driven by the number of ferromagnetic atomic layers.

6. Effect of Ta thickness on DMI

For comparison and to verify that the DMI energies calculated are robust as a function of Ta thickness, we show in Figure S6 the $E_{DMI}$ as a function of the oxygen position in the Fe/TaOx structures for $Fe_7Ta_3$ (red circles) and $Fe_7Ta_5$ (blue triangles). The values are within good agreement namely in the DMI sign.

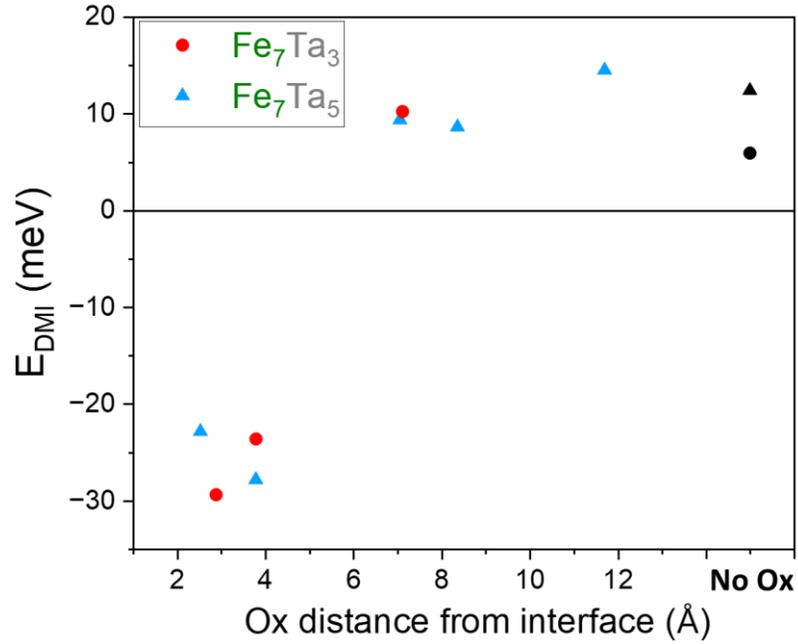

*Figure S6 DMI energy for two Ta thicknesses, with various position of the oxygen in the Ta layers.*

7. <u>Interlayer and intralayer contributions of DMI</u>

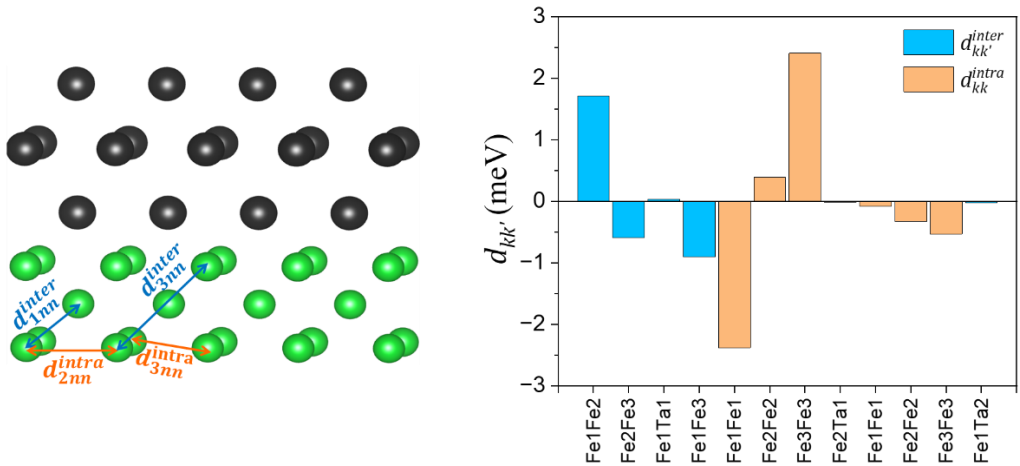

*Figure S7 The intra- and interlayer DMI interactions up to the third nearest neighbor are represented on the crystal structure of Fe/Ta system. The calculated layer-resolved intra- and interlayer contributions to the DMI energy are represented by orange and blue bars, respectively.*

The DMI energy in a bcc structure can be splitted into two contributions, interlayer $E_{DMI}^{inter}$ and intralayer $E_{DMI}^{intra}$. Considering up to third nearest neighbor, we find:

$$E_{DMI}^{inter} = 8\sqrt{2} \sin\theta \, d_{1nn}^{inter} + 8\sqrt{2} \, d_{3nn}^{inter}, \text{ and}$$

$$E_{DMI}^{intra} = 8 \, d_{2nn}^{intra} + 8\sqrt{2} \, d_{3nn}^{intra},$$

where $d$ is the microscopic DMI coefficient describing the interaction between two neighboring atoms as represented on the crystal structure in Figure S7(left). $\theta$ designates the angle between the unit vector connecting the two nearest neighbors atoms in the adjacent planes and the normal to the surface.

To calculate the layer resolved DMI $E_k^{intra}$ of atomic layer $k$ from first-principles, we construct a spin spiral in this layer and fix the spins of the other layers in the $y$-crystallographic direction. Besides, $E_{k,k'}^{inter}$ is calculated by constructing two spirals in the bi-layers $k$ and $k'$. The results of those series of calculations for Fe$_3$Ta$_3$ structure are presented in Figure S7. It can be seen that the largest intralayer DMI contributions are the interfacial Fe1 $E_{Fe1}^{intra}$ and the vacuum interface Fe3 showing opposite signs. On the other hand, the large interlayer DMI $E_{Fe1,Fe2}^{inter}$ is compensated by the opposite sign of $E_{Fe1,Fe3}^{inter}$ together with $E_{Fe2,Fe3}^{inter}$. In an overall picture, the summation of the layer resolved DMI energy comprising both intra- and interlayer contributions $\sum_k E_k^{intra} + \sum_{k,k'} E_{k,k'}^{inter} = -7.84$ meV is within a margin of 6% difference from the calculated total DMI energy $-8.34$ meV. This slight difference, which demonstrates the accuracy of our results, is attributed to the fact that calculations with different constrained configurations cannot be strictly equivalent.

8. Structural relaxation effect with Fe thickness variation

The dependence of the Fe/Ta interfacial distance on the Fe layer thickness is attributed to the sensitivity of the structural relaxations in this ultrathin regime. We show in Figure S8 the interlayer distances of the structures Fe$_{(3-9)}$/Ta$_3$ considered in this study. The Fe-Fe interlayer distance vary notably when increasing the Fe thickness from three to five layers. However, those values become almost stable beyond five layers as the Fe-Fe inner layers converge to the bulk interlayer distance (1.433 Å). In turn, the variation of the Fe-Fe interlayer distance affects the Fe-Ta interface more substantially in ultrathin Fe layers (three to five). This structural relaxation effect was also reported for Fe/MgO interface [2].

In fact, this structural effect is correlated to the layer resolved spin-orbit coupling energy $\Delta E_{soc}^k$ (Figure 5 (a) of the main text). The three Fe atoms close to vacuum have a similar contribution to SOC energy for Fe$_5$Ta$_3$, Fe$_7$Ta$_3$ and Fe$_9$Ta$_3$. The trends in the structural parameters and the SOC energies with the Fe thickness variation highlight the critical thickness limit of five monolayers beyond which the two interfaces can be considered decoupled, while they cannot be disentangled in the ultrathin Fe$_3$Ta$_3$ case.

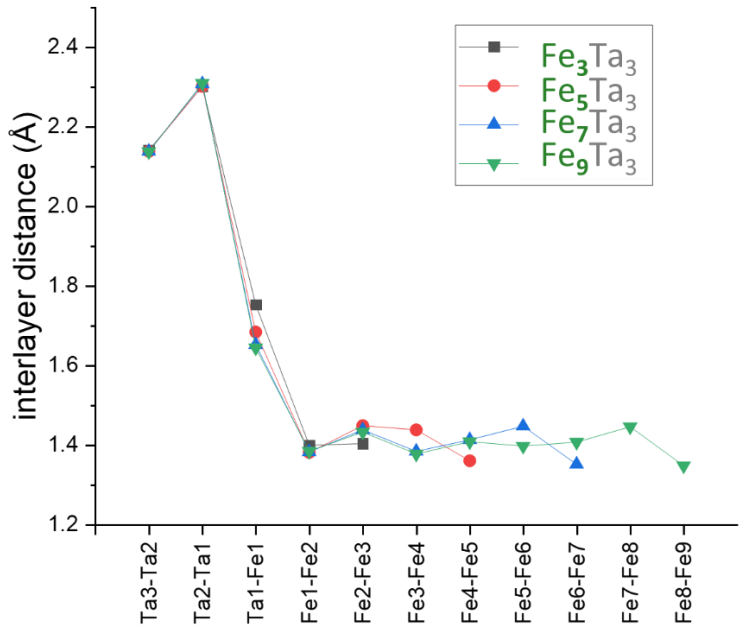

Figure S8 Variation of the interlayer distances with increasing the Fe thickness in structures $Fe_{(3-9)}/Ta_3$